\documentclass[12pt,a4paper,showkeys]{revtex4}
\usepackage{epsfig}
\usepackage{bm}
\usepackage{amsmath}
\usepackage{amssymb}
\usepackage{graphics}
\usepackage[linkcolor=red,citecolor=green,urlcolor=blue]{hyperref}
\usepackage{graphicx,color}
\usepackage{hyperref}

\begin{document}

\title{ \textbf{Associated production of one particle and a Drell-Yan pair in hadronic collisions}}
\author{Federico Alberto Ceccopieri}
\email{federico.alberto.ceccopieri@cern.ch}
\affiliation{IFPA, Universit\'e de Li\`ege,  All\'ee du 6 ao\^ut, B\^at B5a,   \\4000
Li\`ege, Belgium}

\begin{abstract}
We propose a collinear factorization formula 
for the associated production of one particle and a Drell-Yan pair 
in hadronic collisions. It is shown that additional collinear 
singularities appearing in the next-to-leading order calculations that can not be
factorized into parton and fragmentation functions are systematically 
renormalized by introducing fracture functions.
Next-to-leading order coefficient functions for cross-sections double differential 
in the fractional energy of the identified hadron and 
lepton pair invariant mass are presented. 
\end{abstract}
\maketitle

\section{Introduction}
\noindent
The description of particle production in hadronic collisions 
is interesting and challenging in many aspects. Perturbation 
theory can be applied whenever a sufficiently hard scale characterizes
the scattering process. 
The comparison of early LHC charged particle spectra with 
next-to-leading order perturbative QCD predictions~\cite{stratmann}
shows that the theory offers a rather good description of data at 
sufficiently high hadronic transverse momentum, of the order 
of a few GeV. For inelastic scattering processes at even lower transverse
momentum, the theoretical description in terms of perturbative QCD breaks 
down since both the coupling and partonic matrix elements diverge as the transverse momenta 
of final state parton vanish.
In this paper we will study the semi-inclusive version of the Drell-Yan process, 
$ H_1 + H_2 \rightarrow H + \gamma^* + X$, in which one particle is tagged in the final 
state together with the Drell-Yan pair. 
In such a process the high invariant mass of the lepton pair, $Q^2$, constitutes the perturbative
trigger which guarantees the applicability of perturbative QCD.
The detected hadron $H$ could then be used, without any phase space restriction,
as a local probe to investigate particle production mechanisms.
The evaluation of $\mathcal{O}(\alpha_s)$ corrections shows   
that there exists a class of collinear singularities 
escaping the usual renormalization procedure which amounts to reabsorb  collinear divergences 
into a redefinition of bare parton and fragmentation functions.
Such singularities are likely to appear in every fixed order calculation
in the same kinematical limits spoiling the convergence of the perturbative 
series. We therefore show how to improve the theoretical description providing 
a generalized procedure for the factorization of such additional collinear singularities.
Most important, the latter is the same as the one proposed in Deep Inelastic Scattering~\cite{Graudenz}
where the same  collinear singularities pattern is also found, 
confirming the universality of the collinear radiation between different hard processes.  
The latter will make use the concept of fracture functions and the renormalization group equations associated with 
them~\cite{Trentadue_Veneziano}. In a pure parton model approach, 
these non-perturbative distributions effectively describe 
the hadronization of the spectators system in hadron-induced reactions.   
We will demonstrate that the transverse-momentum integrated cross-section is finite and valid for 
all transverse momentum of the detected hadron, without any 
restriction imposed by the singular behaviour of matrix elements. 
We further note that this process is the single-particle counterpart 
of electroweak-boson plus jets associated production~\cite{zjets}, 
presently calculated at nex-to-leading
order accuracy with up to three jets in the final state~\cite{w3jet}.
One virtue of jet requirement is that it indeed avoids 
the introduction of fragmentation functions to model the final state, 
which are instead one of the basic ingredients entering our formalism.
At variance with our case, however, jet reconstructions at very 
low transverse momentum starts to be challenging~\cite{cacciari}
and it makes difficult the study of this interesting portion of the produced particle spectrum.  \\

In Ref.~\cite{newfracture} a first attempt was made to study the associated radiation 
in Drell-Yan type process with the aid of the concept of 
extended fracture functions~\cite{extendedM}.
The collinear factorization formula was then studied in Ref.~\cite{SIDYmy}
where we were mainly concerned with the study of hard diffractive processes in hadronic collisions. 
The latter are a special case of associated production in which the tagged 
hadron is a proton at very low transverse momentum and almost the energy of the incoming proton.   
In this paper we consider the production of a generic hadron and report all the coefficient functions for the cross-sections differential in the invariant mass of the lepton pair and fractional momentum of the detected hadron.\\

The outline of this paper is as follows. In Section~\ref{sec:iDY} 
we set the notation and briefly review the factorization of collinear singularities 
in the inclusive Drell-Yan process. 
In Section~\ref{NLOcentral} we evaluate the 
$\mathcal{O}(\alpha_s)$ contributions to the cross-section in which the hadron is produced 
by the fragmentation of a final state parton. 
In Section~\ref{assDY} we define the parton model cross-section for the associated production 
of one particle and a Drell-Yan pair in term of fracture functions and evaluate the corresponding  
$\mathcal{O}(\alpha_s)$ corrections.
In Section~\ref{NLOfinite} we present the finite cross sections in terms 
of renormalized fracture functions, distribution functions and fragmentation functions.  
The paper closes with a summary and some conclusions. Technical details and explicit formulas 
are collected in the appendix.

\section{Inclusive Drell-Yan process}
\label{sec:iDY}
\noindent
In this section we briefly review the factorization of collinear singularities 
for the inclusive Drell-Yan process, which will be then properly generalized 
when dealing with the associated production case. 
Consider therefore the collision of two hadrons $H_1$ and $H_2$ of momenta 
$P_1$ and $P_2$, respectively:
\begin{equation}
\label{incl_process}
H_1(P_1) + H_2(P_2) \rightarrow \gamma^*(q) +X \,,
\end{equation} 
where $\gamma^*$ stands for the virtual photon of invariant mass $Q^2\gg \Lambda_{QCD}^2$
and $X$ for the unobserved part of the final state.
The dilepton pair detected in the final state is the decay product of 
a virtual photon $\gamma^*$ created, to lowest order, by the annihilation 
of a quark and an antiquark with momenta $p_1=x_1 P_1$ and $p_2= x_2 P_2$, respectively, 
and both assumed to be collinear to their parent hadron.
The corresponding differential cross-section threfore is given by~\cite{DY}
\begin{equation}
\label{iDY}
\frac{d\sigma(\tau)}{dQ^2}=\sigma_0
\int_{\tau}^{1} \frac{dx_1}{x_1} \int^{1}_{\frac{\tau}{x_1}} \frac{dx_2}{x_2} 
\sum_q e_q^2 \Big[ f_q^{[1]}(x_1) f_{\bar{q}}^{[2]}(x_2)+(q \leftrightarrow \bar{q}) \Big] 
\delta(1- w )\,.
\end{equation}
The puntiform cross-section is denoted by $\sigma_0=4\pi\alpha^2_{\makebox{\tiny{em}}} / 9 S_h Q^2$.
The hadronic centre of mass energy is denoted by $S_h=(P_1+P_2)^2$ and the relevant 
ratio $\tau$ is defined by $\tau=Q^2/S_h$.
To zeroth order in $\alpha_s$, 
the partonic sub-process squared energy $\hat{s}=(p_1+p_2)^2$ is fixed  to be
$s=x_1 x_2 S_h=Q^2$, from which the constraint $w=\tau/x_1 x_2=1$ follows. The flavour sum runs on quarks only.
The puntiform cross-sections $\sigma_0$ is
weighted by the convolution over parton distributions, $f_q^{[1]}$ and $f_{\bar{q}}^{[2]}$,
 evaluated at fractional momenta $x_1$ and $x_2$.
The integration limits are fixed by momentum conservation. 
We further assign to each parton distribution 
function an index $n=[1,2]$ denoting the hadron from 
which the quark (or antiquark) has been extracted.
Calculations are performed in dimensional regularization with space-time dimension set to $n=4-2\epsilon$. 
Following Ref.~\cite{DYNLO}, 
the cross-section for the partonic sub-process $q(p_1)+\bar{q}(p_2)\rightarrow \gamma^*(q)$ 
is given by
\begin{equation}
\label{LO}
\frac{d\sigma_{q\bar{q}}(w,Q^2)}{dQ^2}=\delta(1-w)\,.
\end{equation}
It defines the normalization of the partonic Drell Yan cross section and corresponds to multiplying 
all matrix elements by a factor $2N_c/(1-\epsilon)$, $N_c$ being the 
number of colours. The latter factor  
is already absorbed  in the puntiform cross-section $\sigma_0$ in parton model formula, eq.~(\ref{iDY}). 
The evaluation on next-to-leading corrections is performed by computing the relavant 
real emissions and virtual diagrams. We refer to reader to Ref.~\cite{DYNLO} for 
the explicit expressions of matrix elements and further calculational details.
Both real and virtual terms develop poles in $\epsilon^{-2}$ which mutually cancel
when these contributions are added so that one finally obtains
\begin{eqnarray}
\label{iDY_NLO_sing}
&&\hspace{-1.5cm}\frac{d\sigma(\tau)}{dQ^2}=\sigma_0
\int_{\tau}^{1} \frac{dx_1}{x_1} \int_{\frac{\tau}{x_1}}^{1} \frac{dx_2}{x_2} 
\sum_q e_q^2 \Bigg\{  \\
&&  \Big[ f_q^{[1]}(x_1) f_{\bar{q}}^{[2]}(x_2) +(q \leftrightarrow \bar{q})\Big]
\Big[ \delta(1-w)
-\frac{2}{\epsilon}\frac{\alpha_s}{2\pi} P_{qq}(w) \, c_0 +\frac{\alpha_s}{2\pi}
\widetilde{C}_{q\bar{q}}(w) \Big]+\nonumber\\
&& \hspace{3.5cm} + \Big[ \big(f_g^{[1]}(x_1) f_{\bar{q}}^{[2]}(x_2)+f_q^{[1]}(x_1) f_g^{[2]}(x_2)
\big) 
+(q \leftrightarrow \bar{q})\Big] \cdot \nonumber\\
&& \hspace{7cm} \cdot \Big[-\frac{1}{\epsilon}\frac{\alpha_s}{2\pi}
P_{qg}(w) \, c_0 +\frac{\alpha_s}{2\pi}
\widetilde{C}_{qg}(w) \Big]\Bigg\}\,.\nonumber
\end{eqnarray} 
In the previous equation collinear singularities appear as poles in $\epsilon^{-1}$
multiplying the leading order~\cite{DGLAP} splitting functions $P_{ij}(w)$.
The adimensional factor $c_0$ appearing in eq.~(\ref{iDY_NLO_sing}) reads 
\begin{equation}
\label{c0}
c_0=\Big( \frac{4 \pi \mu_R^2}{Q^2} \Big)^{\epsilon} 
\frac{\Gamma(1-\epsilon)}{\Gamma(1-2\epsilon)}\,
\end{equation}
where $\mu_R^2$ indicates the renormalization scale. 
The subtraction of singular terms in the partonic cross-sections is performed 
by absorbing the collinear divergences into bare parton distributions,
which in the $\overline{\mbox{MS}}$ scheme amounts to the following redefinition: 
\begin{equation}
\label{f_renorm}
f_i(x)=\int_{x}^{1}\frac{du}{u} \Big[ \delta_{ij}\delta(1-u) +\frac{1}{\epsilon}\frac{\alpha_s}{2\pi}
\frac{\Gamma(1-\epsilon)}{\Gamma(1-2\epsilon)}
\Big( \frac{4 \pi \mu_R^2}{\mu_F^2} \Big)^{\epsilon} P_{ij}(u) \Big] f_j\Big(\frac{x}{u},\mu_F^2\Big)\,.
\end{equation}
The renormalized distributions do depend on the scale, $\mu_F^2$, at which the factorization is performed, 
and their variation with respect to it gives 
governed by DGLAP evolution equations~\cite{DGLAP}. Inserting eq.~(\ref{f_renorm}) into eq.~(\ref{iDY_NLO_sing})
one can explicitely check that collinear singularities, proportional to poles in $\epsilon$, do cancel.
The finite result reads:
\begin{eqnarray}
\label{iDY_NLO_reg}
&&\frac{d\sigma(\tau)}{dQ^2}=\sigma_0
\int_{\tau}^{1} \frac{dx_1}{x_1} \int_{\frac{\tau}{x_1}}^{1} \frac{dx_2}{x_2} 
\sum_q e_q^2 \Bigg\{  \\
&&  \Big[ f_q^{[1]}(x_1,\mu_F^2) f_{\bar{q}}^{[2]}(x_2,\mu_F^2) +(q \leftrightarrow \bar{q})\Big]
\Big[ \delta(1-w) +\frac{\alpha_s}{2\pi}
C_{q\bar{q}}(w,\mu_F^2/Q^2) \Big]+\nonumber\\
&& \Big[ \big(f_g^{[1]}(x_1,\mu_F^2) f_{\bar{q}}^{[2]}(x_2,\mu_F^2)+f_q^{[1]}(x_1,\mu_F^2) 
f_g^{[2]}(x_2,\mu_F^2)
\big) 
+(q \leftrightarrow \bar{q})\Big] \frac{\alpha_s}{2\pi}
C_{qg}(w,\mu_F^2/Q^2) \Bigg\}\,.\nonumber
\end{eqnarray} 
The infrared-finite coefficient functions $\widetilde{C}_{ij}(w)$ 
and $C_{ij}(w)$, defined in the $\overline{\mbox{MS}}$ scheme,
are reported in appendix~\ref{finite_coef}. 
In phenomenological applications it is costumary to set $\mu_F^2=Q^2$ in eq.~(\ref{iDY_NLO_reg}).
This choice removes large logarithms of the ratio $\mu_F^2/Q^2$ appearing 
in the coefficient functions. At the same time, the evaluation of parton distributions 
at a scale $\mu_F^2=Q^2$ accounts for the resummation of such logarithms
via DGLAP evolution equations.

\section{Associated production}
\label{NLOcentral}
\noindent
In the present and next section we consider the next-to-simple generalization 
of reaction in eq.~(\ref{incl_process}), namely the associated production of a Drell-Yan pair 
and an identified hadron $H$ of momentum $h$:
\begin{equation}
\label{ass_process}
H_1(P_1) + H_2(P_2) \rightarrow \gamma^*(q) + H(h) + X\,.
\end{equation}
In particular we will evaluate $\mathcal{O}(\alpha_s)$ corrections to the cross-section double differential 
in the invariant mass $Q^2$ of the lepton
pair and the fractional energy of the identified hadron $H$, 
defined in analogy with $e^+e^-$ annihilation process:
\begin{equation}
\label{zdef}
z=\frac{2h \cdot (P_1+P_2)}{(P_1+P_2)^2}=\frac{2E_H}{\sqrt{S_h}}\,.
\end{equation}
The last equality in the previous equations holds in the hadronic centre of mass 
frame. In this frame $z$ is the proportional to the detected hadron energy, $E_H$, scaled down 
by the beam energy, $\sqrt{S_h}/2$. 
In this section we consider the $\mathcal{O}(\alpha_s)$ production mechanism in which 
the observed hadron $H$ is given by the fragmentation of a, real, final state parton
and address such contribution as \textsl{central}. 
We label the momenta in the partonic sub-process as $i(p_1)+j(p_2) \rightarrow l(k)+\gamma^*(q)$,
where $k$ and $q$ are the four-momenta of the outgoing parton and virtual photon, respectively. 
In general such a correction is expected in the form~\cite{aversa}
\begin{equation}
\label{form}
\frac{d\sigma^{C}(\tau,z,Q^2)}{dQ^2 dz}\propto
\sum_{ijl} \int \frac{dx_1}{x_1} \int \frac{dx_2}{x_2} \int \frac{d\rho}{\rho}
f_i(x_1) f_j(x_2) D_l^H(z/\rho)
\frac{d\hat{\sigma}^{ij\rightarrow l \gamma^*}}{dQ^2 d\rho}\,, 
\end{equation}
where $\rho=2E_l/\sqrt{S_h}$ is the 
fractional energy of the outgoing parton evaluated in the hadronic center of mass frame
and it is the partonic analogue of the hadronic $z$. The sum runs over all possible partonic 
sub-processes
\begin{figure}
\begin{center}
\includegraphics[width=7cm,height=5cm,angle=0]{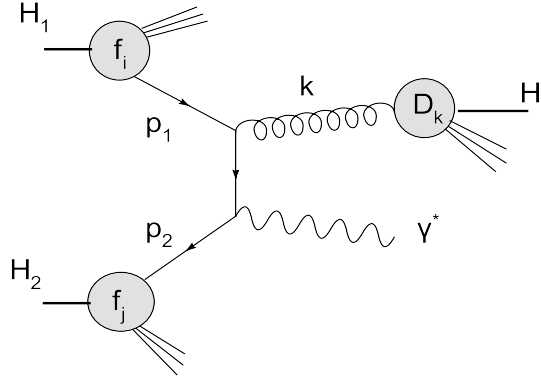}
\caption{One of the diagrams contributing to eq.~(\ref{form}). The observed hadron $H$ is produced  
by the fragmentation of the final state parton $l$.}
\label{fig1}
\end{center}
\end{figure}
and the partonic cross-sections are indicated with $d\hat{\sigma}^{ij\rightarrow l\gamma^*}$.
With respect to eq.~(\ref{iDY}), eq.~(\ref{form}) does contain an additional convolution 
on the fractional energy $\rho$ of the final state parton weighted by the 
fragmentation functions $D_l^H$. The latter gives the probability that a parton with fractional 
energy $\rho$ fragments into the observed hadron $H$ with fractional energy $z$.
One of the diagrams contributing to eq.~(\ref{form}) is depicted in Fig.~(\ref{fig1}).
We found useful to rewrite the convolution formula as a function 
of $y=(1-\cos\theta)/2$, where $\theta$ is its angle between the parton $l$ and the hadron $H_1$
in the hadronic center of mass frame. Within these definitions 
the parton-level invariants $\hat{u}$ and $\hat{t}$ in the matrix elements can be rewritten 
in this frame as 
\begin{eqnarray}
\label{ut}
\hat{t}\;=\;(p_1-k)^2&=& -\; \hat{s}\,(1-w)\;\frac{x_1 \, y}{x_1y+x_2(1-y)}\,,\\
\hat{u}\;=\;(p_2-k)^2&=& -\; \hat{s}\,(1-w)\;\frac{x_2 \, (1-y)}{x_1y+x_2(1-y)}\,,\nonumber
\end{eqnarray}
where $\hat{s}=(p_1+p_2)^2$ and $w=\tau/(x_1x_2)$. The phase space reads:
\begin{equation}
dPS^{(2)}=\frac{1}{8\pi} \Big( \frac{4\pi}{Q^2} \Big)^{\epsilon}\;
\big(x_1 x_2 -\tau\big)^{1-2\epsilon} \; \big(x_1 y +x_2(1-y)\big)^{2\epsilon-2}
\; \frac{\tau^{\epsilon}}{\Gamma(1-\epsilon)} \; dy \; y^{-\epsilon} \; (1-y)^{-\epsilon}\,.
\end{equation}
The angular variable $y$ and the fractional energy $\rho$ of the emitted parton 
are not independent and constrained by: 
\begin{equation}
\label{rho}
\rho(y)=\frac{x_1 x_2-\tau}{x_1 y+ x_2(1-y)}\,.
\end{equation}
The available phase space must take into account that 
there should be enough energy for the production both of the hadron $H$ and 
the virtual photon $\gamma^*$. This phase space constraints will appear 
in the convolution limits in eq.~(\ref{form}). In order to obtain them, 
we notice that the parent parton of the observed hadron is required to have 
a fractional energy  $\rho\ge z > 0$. Applying this last constraint 
to eq.~(\ref{rho}), one is able to determine the
boundaries $r_1$ and $r_2$ in the $x_1$ and $x_2$ convolutions integrals. 
They are both $z$ and $y$ dependent and read
\begin{eqnarray}
\label{r12}
r_1(\tau,z;y)=\frac{\tau+z(1-y)}{1-zy}\,,\nonumber\\
r_2(\tau,z;y)=\frac{\tau+x_1 z y}{x_1-z(1-y)}\,.
\end{eqnarray}
The $\mathcal{O}(\alpha_s)$ corrections in the \textsl{central} region therefore reads
\begin{eqnarray}
\label{DYcentral}
&&\hspace{-0.5cm}\frac{d\sigma^{C}(\tau)}{dQ^2 dz}=\sigma_0
\int_0^1 dy \int_{r_1}^{1} \frac{dx_1}{x_1} \int_{r_2}^{1} \frac{dx_2}{x_2} \sum_q e_q^2 \Big\{ \\
&&\hspace{-0.5cm}
\Big[f_{q}^{[1]}(x_1) \, f_{\bar{q}}^{[2]}(x_2)+(q \leftrightarrow \bar{q})\Big] 
\, D_g^H(z/\rho)
\Big[ -\frac{1}{\rho} \frac{\alpha_s}{2\pi} \frac{c_0}{\epsilon} \widehat{P}_{qq}(w)
[\delta(y)+\delta(1-y)]+\frac{\alpha_s}{2\pi} \widetilde{K}_{q\bar{q}} (y) \Big]+ \nonumber\\
&&\hspace{-0.5cm}
f_{g}^{[1]}(x_1) \Big[f_{\bar{q}}^{[2]}(x_2)
D_{\bar{q}}^H(z/\rho)  + ( q \leftrightarrow \bar{q})\Big] 
\Big[ -\frac{1}{\rho} \frac{\alpha_s}{2\pi} \frac{c_0}{\epsilon} \widehat{P}_{qg}(w) \delta(y)
 +\frac{\alpha_s}{2\pi} \widetilde{K}_{qg} (y)\Big]+ \nonumber\\
&&\hspace{-0.5cm}
f_{g}^{[2]}(x_2) \Big[f_{\bar{q}}^{[1]}(x_1)
D_{\bar{q}}^H(z/\rho) + ( q \leftrightarrow \bar{q})\Big] 
\Big[-\frac{1}{\rho} \frac{\alpha_s}{2\pi} \frac{c_0}{\epsilon} \widehat{P}_{qg}(w) \delta(1-y)
 +\frac{\alpha_s}{2\pi} \widetilde{K}_{qg} (1-y)\Big]\Big\} \,,\nonumber
\end{eqnarray}
where $\rho=\rho(y)$ via eq.~(\ref{rho}) and we 
used the shorthand $\widetilde{K}_{ij} (y)$ for $\widetilde{K}_{ij} (y,x_1,x_2,w)$. 
The expressions for infrared finite coefficients $\widetilde{K}_{ij}$ are collected in appendix~\ref{finite_coef}. 
Eq.~(\ref{DYcentral}) displays two disjoint singular limits for $y\rightarrow 0$ and  $y\rightarrow 1$.  
In order to expose the collinear singularites we have performed an $\epsilon$-expansion on the angular variable $y$:
\begin{eqnarray}
\label{yplus}
(1-y)^{-1-\epsilon}&\simeq& -\frac{1}{\epsilon}\,\delta(1-y)
+\Big(\frac{1}{1-y}\Big)_{+[0,\underline{1}]}-\epsilon\Big( \frac{\ln (1-y)}{1-y}\Big)_{+[0,\underline{1}]} 
 +\mathcal{O}(\epsilon^2) \,, \\
y^{-1-\epsilon}&\simeq& -\frac{1}{\epsilon}\,\delta(y)
+\Big(\frac{1}{y}\Big)_{+[\underline{0},1]}-\epsilon\Big( \frac{\ln y}{y}\Big)_{+[\underline{0},1]} 
+ \mathcal{O}(\epsilon^2) \,. 
\end{eqnarray}
The unregularized splitting functions $\widehat{P}_{ij}$ appearing in eq.~(\ref{DYcentral}) 
are given by~\cite{KUV}
\begin{equation}
\widehat{P}_{qq}(w)=C_F \frac{1+w^2}{1-w}, \;\;\; \widehat{P}_{qg}(w)=T_R [w^2+(1-w)^2]\,,
\end{equation}
with $C_F=4/3$ and $T_R=1/2$.
We wish to conclude this section by noting that the collinear divergences 
appearing in eq.~(\ref{DYcentral}) do correspond to configurations in which 
the parent parton of the observed hadron is collinear to the incoming parton. 
Such divergences at vaninishing transverse momentum escape, as shown 
already in the context of Semi-Inclusive Deep Inelastic Scattering~\cite{Graudenz},  
any factorization in terms of renormalized parton distributions and fragmentation functions. 
While for many practical applications they are regularized introducing an arbitrary cut-off on
the produced hadron transverse momentum, configurations which give rise to these divergences 
will be present at every order in perturbative calculations. 
Fracture functions together with their own renormalization group equations 
can be shown to provide the correct tool to perform the resummation to all orders of large logarithmic contributions
coming from the factorization of such collinear singularities.

\section{Corrections in the target region}
\label{assDY}
\noindent

\begin{figure}[t]
\begin{center}
\includegraphics[width=5cm,height=5cm,angle=0]{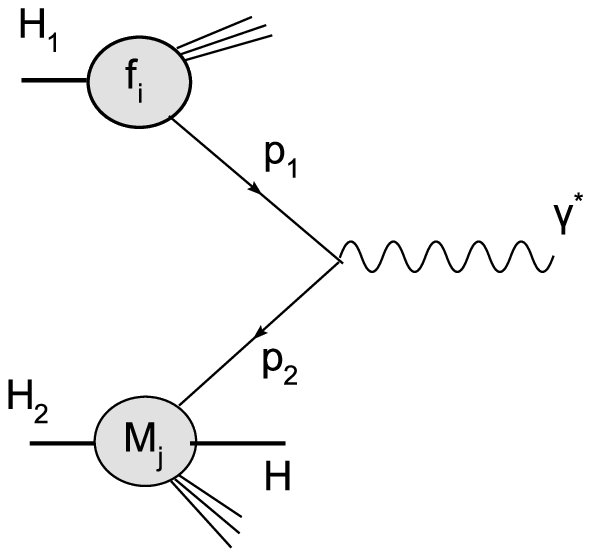}
\includegraphics[width=5cm,height=5cm,angle=0]{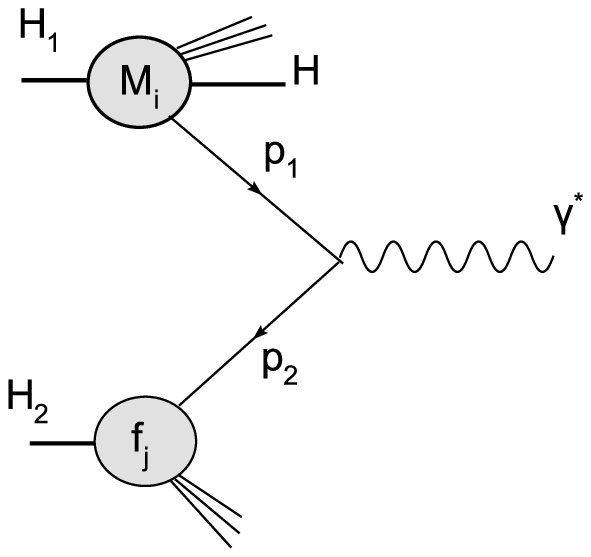}
\caption{Pictorial representation of the parton model formula in eq.~(\ref{siDY}). The hadron $H$ is non-perturbatively  produced by a fracture functions.}
\label{fig56}
\end{center}
\end{figure}

\begin{figure}
\begin{center}
\includegraphics[width=6cm,height=5cm,angle=0]{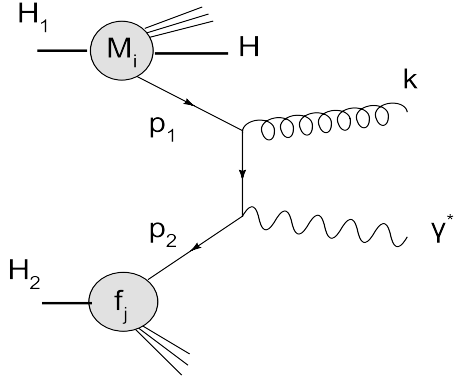}
\caption{Example of $\mathcal{O}(\alpha_s)$ real corrections contributing to eq.~(\ref{siDY_target}). 
The hadron $H$ is non perturbatively produced by a fracture function.}
\label{fig7}
\end{center}
\end{figure}

To lowest order in the QCD coupling no hadron can be produced in the final state
since QCD radiation is absent. In this case we assume that hadron production 
is described by fracture functions $M^{H/H_1}_{i}(x_1,z)$ and $M^{H/H_2}_{j}(x_2,z)$.
These non-perturbative distributions give the conditional probability of finding 
a parton $i$($j$) of fractional momemntum $x_1$($x_2$) in the incoming hadron $H_1$($H_2$) 
while an hadron $H$, with fractional momentum $z$, is detected in the final state~\cite{Trentadue_Veneziano}.
In a pure parton model approach they describe hadron production 
in the target fragmentation region of $H_1$  or $H_2$.
The latter regions, denoted by $\mathcal{R}_{T_1}$ and  $\mathcal{R}_{T_2}$, respectively, 
can be defined as  $\theta=0$ and 
$\theta=\pi$, where $\theta$
is the angle between $H$ and $H_1$ defined in the centre of mass frame.
Fracture function were originally introduced to describe  
hadron production in the Deep Inelastic Scattering target fragmentation
region.
A renormalization group evolution equations were derivered~\cite{Trentadue_Veneziano}
with the aid of Jet Calculus technique~\cite{KUV}.
Subsequently the soft and collinear factorization of these distributions in Semi-Inclusive DIS 
was proven respectively in Refs.~\cite{Fact_M_soft,Fact_M_coll}. 
A complete one loop calculation was presented in Ref.~\cite{Graudenz}, confirming
the factorization conjecture first formulated in Ref.~\cite{Trentadue_Veneziano}.
We emphasize that by virtue of the factorization theorem, fracture functions 
are univeral distributions, at least in the context of Semi-Inclusive DIS.
In hadronic collisions however such a proof does not exist and counter example to it 
have been given in the context of diffractive production in Ref.~\cite{Fact_M_soft}.
The main motivation for using fracture functions in the present context is that, 
as we shall prove in the following sections, they allow us to systematically factorize  
collinear divergences occuring in the evaluation of the partonic cross-sections. 
Given these assumptions, we now present the parton-model formula, depicted in Fig.~(\ref{fig56}), for the associated 
production case: 
\begin{eqnarray}
\label{siDY}
&&\hspace{-0.7cm}\frac{d\sigma^{H}(\tau)}{dQ^2 dz}=\\
&&\sigma_0 
\int_{\tau}^{1-z} \frac{dx_1}{x_1} \int_{\frac{\tau}{x_1}}^{1} \frac{dx_2}{x_2} 
\sum_q e_q^2 \Big[ M_q^{[1]}(x_1,z) f_{\bar{q}}^{[2]}(x_2) + M_{\bar{q}}^{[1]}(x_1,z) f_q^{[2]}(x_2)\Big]\,
\delta( 1-w)+\nonumber\\
&&\sigma_0 \int_{\frac{\tau}{1-z}}^{1} \frac{dx_1}{x_1} \int_{\frac{\tau}{x_1}}^{1-z} \frac{dx_2}{x_2} 
\sum_q e_q^2 \Big[ M_q^{[2]}(x_2,z) f_{\bar{q}}^{[1]}(x_1) + M_{\bar{q}}^{[2]}(x_2,z) f_q^{[1]}(x_1)\Big]\,
\delta( 1-w)\,.\nonumber
\end{eqnarray} 
The integration limits of convolution integrals in both lines of eq.~(\ref{siDY}) 
are given by momentum conservation:
\begin{eqnarray}
1-z\ge x_1 x_2 \ge \tau \;\; \makebox{and} \;\; x_1+z \le 1 \;\; \makebox{in} \;\;\mathcal{R}_{T_1} \,,\\
1-z\ge x_1 x_2 \ge \tau \;\; \makebox{and} \;\; x_2+z \le 1 \;\; \makebox{in} \;\;\mathcal{R}_{T_2}\,.
\end{eqnarray}  
Phase space integrations are asymmetric since each fracture function selects its 
own fragmentation region. To uniform the notation, we exchange the superscript $H/H_1$ ($H/H_2$) for  
$[1]$ ($[2]$) which proves to be useful for the bookkeeping of the various contributions.
Fracture functions in eq.~(\ref{siDY}) are normalized according to the constraint
\begin{equation}
\label{sum_rule}
\sum_{H} \int dz \frac{d\sigma^{H}(\tau)}{dQ^2 dz} = \frac{d\sigma(\tau)}{dQ^2}\,.
\end{equation}
The above constraint must me fullfilled irrespective to the order of the perturbative 
calculations. It is interesting to note that for the inclusive Drell-Yan
cross-sections appearing on the right hand side of eq.~(\ref{sum_rule}), 
the factorization theorem guarantees that the corresponding cross-sections 
can be described by universal parton distributions functions.
On the left hand side of eq.~(\ref{sum_rule}), we have instead no guarantee 
that fracture functions eventually extracted from Deep Inelastic Scattering can be succesfully used
in hadronic collisions. Eq.~(\ref{siDY}) is therefore both a factorization and 
a cross-section \textsl{conjecture} for the process under examination.
In the remainder of this section we consider the evaluation of $\mathcal{O}(\alpha_s)$
corrections to the parton model formula, eq.~(\ref{siDY}). 
We address it as \textsl{target} contributions to distinguish them from the one evaluated in Sec.~\ref{NLOcentral}.
As already stated, when the final state hadron is observed in  
$\mathcal{R}_{T_1}$ or $\mathcal{R}_{T_2}$, we assume that it has been 
non-perturbatively produced  from a fracture functions. Final state partons occurring
in $\mathcal{O}(\alpha_s)$ corrections to eq.~(\ref{siDY}) must be therefore integrated over phase space 
and virtual corrections added. A diagram contributing to eq.~(\ref{siDY_target}) is depicted in 
Fig.~(\ref{fig7}). 
The calculation closely follows the one already presented in Sec.~\ref{sec:iDY}
for the inclusive Drell-Yan process upon the exchange of a parton distributions with a fracture functions and 
taking into account momentum conservation in integrations limits.
We therefore just quote the final result, valid up to $\mathcal{O}(\alpha_s)$:
\begin{eqnarray}
\label{siDY_target}
&&\hspace{-1cm}\frac{d\sigma^{T}(\tau)}{dQ^2 dz}=\sigma_0
\int_{\tau}^{1-z} \frac{dx_1}{x_1} \int_{\frac{\tau}{x_1}}^{1} \frac{dx_2}{x_2} 
\sum_q e_q^2 \Bigg\{  \\
&& \hspace{1cm} \Big[ M_q^{[1]}(x_1,z) f_{\bar{q}}^{[2]}(x_2) +(q \leftrightarrow \bar{q})\Big]
\Big[ \delta(1-w)
-\frac{2}{\epsilon}\frac{\alpha_s}{2\pi} P_{qq}(w) \, c_0 +\frac{\alpha_s}{2\pi}
\widetilde{C}_{q\bar{q}}(w) \Big]+\nonumber\\
&& \hspace{1cm} \Big[ \big(M_g^{[1]}(x_1,z) f_{\bar{q}}^{[2]}(x_2)+M_q^{[1]}(x_1,z) f_g^{[2]}(x_2)
\big) 
+(q \leftrightarrow \bar{q})\Big]\cdot \nonumber\\
&&\hspace{8cm}\cdot\Big[-\frac{1}{\epsilon}\frac{\alpha_s}{2\pi}
P_{qg}(w) \, c_0 +\frac{\alpha_s}{2\pi}
\widetilde{C}_{qg}(w) \Big]\Bigg\}+\nonumber\\
&& \hspace{0.7cm}+\sigma_0\int_{\frac{\tau}{1-z}}^1 \frac{dx_1}{x_1} \int_{\frac{\tau}{x_1}}^{1-z} \frac{dx_2}{x_2} 
\sum_q e_q^2 \Bigg\{ \nonumber\\
&& \hspace{1cm} \Big[ M_q^{[2]}(x_2,z) f_{\bar{q}}^{[1]}(x_1) +(q \leftrightarrow \bar{q})\Big]
\Big[ \delta(1-w)
-\frac{2}{\epsilon}\frac{\alpha_s}{2\pi} P_{qq}(w) \, c_0 +\frac{\alpha_s}{2\pi}
\widetilde{C}_{q\bar{q}}(w) \Big]+\nonumber\\
&& \hspace{1cm} \Big[ \big(M_g^{[2]}(x_2,z) f_{\bar{q}}^{[1]}(x_1)+M_q^{[2]}(x_2,z) f_g^{[1]}(x_1)
\big) 
+(q \leftrightarrow \bar{q})\Big]\cdot \nonumber\\
&& \hspace{8cm} \cdot\Big[-\frac{1}{\epsilon}\frac{\alpha_s}{2\pi}
P_{qg}(w) \, c_0 +\frac{\alpha_s}{2\pi}
\widetilde{C}_{qg}(w) \Big]\Bigg\}\,. \nonumber
\end{eqnarray} 

\noindent
Comparing eq.~(\ref{iDY_NLO_sing}) with eq.~(\ref{siDY_target}) reveals that in target fragmentation region 
the structure of collinear singularities is the same as in the inclusive Drell-Yan case, as expected. 
The main change is just a restriction on phase space integrals since the production of the 
Drell-Yan pair of a given invariant mass $Q^2$ must be, by energy-momentum conservation,
compatible with the observation of a hadron in the final state
with fractional momentum $z$. 

\section{Finite cross-sections at NLO}
\label{NLOfinite}
\noindent
In this section we will describe the collinear factorization procedure which
must be applied in order to get infrared finite results for the cross-section under examination. 
We have already noted, by comparing eq.~(\ref{iDY_NLO_sing}) and eq.~(\ref{siDY_target}), 
that the structure of collinear singularities in the target fragmentation
region is identical to the one found in the inclusive Drell-Yan case.  
We may expect that renormalized fracture functions are defined in a way similar to that of 
renormalized parton densities in eq.~(\ref{f_renorm}).
As it was firstly obtained in the original analysis of Ref.~\cite{Trentadue_Veneziano}
and confirmed in the one loop calculation of Ref.~\cite{Graudenz}, 
the renormalized fracture functions obey a somewhat more involved subtraction with 
respect to parton distributions. In the $\overline{\textrm{MS}}$ scheme the redefinition
of bare fracture functions reads:
\begin{multline}
\label{M_renorm}
M_i^{H/H_1}(x,z)=\int_{x/(1-z)}^{1}\frac{du}{u} \Big[ \delta_{ij}\delta(1-u) 
+\frac{1}{\epsilon}\frac{\alpha_s}{2\pi}
\frac{\Gamma(1-\epsilon)}{\Gamma(1-2\epsilon)}
\Big( \frac{4 \pi \mu_R^2}{\mu_F^2} \Big)^{\epsilon} P_{ij}(u) \Big] 
M_j^{H/H_1}\Big(\frac{x}{u},z,\mu_F^2\Big)\,+\\
+\int_x^{x/(x+z)} \frac{du}{u} \frac{1}{1-u}\frac{u}{x}
\frac{1}{\epsilon}\frac{\alpha_s}{2\pi}
\frac{\Gamma(1-\epsilon)}{\Gamma(1-2\epsilon)}
\Big( \frac{4 \pi \mu_R^2}{\mu_F^2} \Big)^{\epsilon}
\widehat{P}_{ij}(u)\,f_{j/H_1}\Big(\frac{x}{u}\Big)\, D_l^H\Big( \frac{z u}{x(1-u)}\Big)\,,
\end{multline}
where in our notation $\widehat{P}_{ij}(u) = \widehat{P}_{(l)i\leftarrow j}(u)$. 
The first term on r.h.s of eq.~(\ref{M_renorm}) has the same subtraction structure as 
for parton distribution, eq.~(\ref{f_renorm}). The singularity is due to collinear 
radiation accompaining the active parton, while the hadron in the final state 
is non perturbatively produced by the fracture functions itself.  
In the second term of eq.~(\ref{M_renorm}) the singularity is due configurations in which  
the parent parton of the observed  hadron is collinear to the incoming parton.
The factorization procedure 
is accounted for by substituting in eq.~(\ref{siDY_target}) the bare
fracture and distributions functions by their renormalized version in eq.~(\ref{f_renorm})
and eq.~(\ref{M_renorm}). Renormalized parton distributions and fracture functions 
homogeneous terms do cancel all singularities present in eq.~(\ref{siDY_target}).
The additional singularities in eq.~(\ref{DYcentral}) are cancelled by the combination 
of parton distributions and fracture functions inhomogeneous renormalization terms.
The final result, up to order $\mathcal{O}(\alpha_s)$, is obtained adding the 
the various contributions:
{\setlength\arraycolsep{0pt}
\begin{eqnarray}
\label{siDY_total_finite}
&&\frac{d\sigma^{H}(\tau)}{dQ^2 dz}=\frac{d\sigma^{H,T}(\tau)}{dQ^2 dz}\;+ \;
\frac{d\sigma^{H,C}(\tau)}{dQ^2 dz}=\sigma_0 \sum_q e_q^2 \Bigg\{ \\
&&
\int_{\tau}^{1-z} \frac{dx_1}{x_1} \int_{\frac{\tau}{x_1}}^{1} \frac{dx_2}{x_2} 
\bigg\{ \Big[ M_q^{[1]}(x_1,z,\mu_F^2) f_{\bar{q}}^{[2]}(x_2,\mu_F^2) +(q \leftrightarrow \bar{q})\Big]
\Big[ \delta(1-w)+\frac{\alpha_s}{2\pi} C_{q\bar{q}}\Big( w,\frac{\mu_F^2}{Q^2}\Big) \Big]+\nonumber\\
&& \hspace{1.0cm}  \Big[ \big(M_g^{[1]}(x_1,z,\mu_F^2) f_{\bar{q}}^{[2]}(x_2,\mu_F^2)+M_q^{[1]}(x_1,z,\mu_F^2) 
f_g^{[2]}(x_2,\mu_F^2)
\big) +(q \leftrightarrow \bar{q})\Big] \frac{\alpha_s}{2\pi}
C_{qg}\Big( w, \frac{\mu_F^2}{Q^2}\Big) \bigg\}+\nonumber\\
&& \hspace{0cm}\int_{\frac{\tau}{1-z}}^1 \frac{dx_1}{x_1} \int_{\frac{\tau}{x_1}}^{1-z} \frac{dx_2}{x_2} 
\bigg\{ \Big[ M_q^{[2]}(x_2,z,\mu_F^2) f_{\bar{q}}^{[1]}(x_1,\mu_F^2) 
+(q \leftrightarrow \bar{q})\Big] \Big[ \delta(1-w)+\frac{\alpha_s}{2\pi}
C_{q\bar{q}}\Big( w,\frac{\mu_F^2}{Q^2}\Big) \Big]+\nonumber\\
&& \hspace{1.0cm}  \Big[ \big(M_g^{[2]}(x_2,z,\mu_F^2) f_{\bar{q}}^{[1]}(x_1,\mu_F^2)
+M_q^{[2]}(x_2,z,\mu_F^2) f_g^{[1]}(x_1,\mu_F^2)\big) +(q \leftrightarrow \bar{q})\Big] \frac{\alpha_s}{2\pi}
C_{qg}\Big( w, \frac{\mu_F^2}{Q^2}\Big) \bigg\}+\nonumber\\
&& \hspace{0.0cm} \int_0^1 dy \int_{r_1}^{1} \frac{dx_1}{x_1} \int_{r_2}^{1} \frac{dx_2}{x_2} 
\bigg\{ \Big[f_{q}^{[1]}(x_1,\mu_F^2) \, f_{\bar{q}}^{[2]}(x_2,\mu_F^2)+(q \leftrightarrow \bar{q})\Big] 
\, D_g^H\big(z/\rho,\mu_F^2\big) \frac{\alpha_s}{2\pi} K_{q\bar{q}}
\Big(y,\frac{\mu_F^2}{Q^2}\Big)+ \nonumber\\
&&\hspace{3.5cm}
 f_{g}^{[1]}(x_1,\mu_F^2) \Big[f_{\bar{q}}^{[2]}(x_2,\mu_F^2)
D_{\bar{q}}^H\big(z/\rho,\mu_F^2\big)  + ( q \leftrightarrow \bar{q})\Big] 
\, \frac{\alpha_s}{2\pi} K_{qg}
\Big(y,\frac{\mu_F^2}{Q^2}\Big)+ \nonumber\\
&&\hspace{3.5cm}
 f_{g}^{[2]}(x_2,\mu_F^2) \Big[f_{\bar{q}}^{[1]}(x_1,\mu_F^2)
D_{\bar{q}}^H\big(z/\rho,\mu_F^2\big)  + ( q \leftrightarrow \bar{q})\Big] 
\, \frac{\alpha_s}{2\pi} K_{qg}
\Big(1-y,\frac{\mu_F^2}{Q^2}\Big)\bigg\}\Bigg\}\,.\nonumber
\end{eqnarray}
We have used the shorthand $K_{ij}(y,\mu_F^2/Q^2)$ for $K_{ij}(y,x_1,x_2,w,\mu_F^2/Q^2)$.
The explicit form of the finite coefficient functions $K_{q\bar{q}}$ and $K_{qg}$ 
is reported in appendix~\ref{finite_coef}.
In the last three lines of eq.~(\ref{siDY_total_finite})
we let depend parton distributions and fragmentation functions on the factorization scale
since this replacement induces subleading corrections to the current accuracy.

\section{Summary and Conclusions}
\label{summary}
\noindent
In this paper we have calculated the $\mathcal{O}(\alpha_s)$ corrections to the associated 
production of one particle and a Drell-Yan pair. 
Additional collinear singularities found in the perturbative calculations
do correspond to configurations in which the parent parton of the observed hadron 
is collinear to the incoming parton. These singularities can, to $\mathcal{O}(\alpha_s)$, 
be consistently absorbed into renormalized fracture functions and resummed to all orders
by using the evolution equation given in Refs.~\cite{Trentadue_Veneziano, Graudenz}.
With this technique, the presented cross-sections does not require 
any cut in the transverse momentum of the observed particle, while a perturbative 
treatment is guarantee by the presence of high invariant mass dilepton pair.  
Quite imprtantly, the factorization of collinear singularities in the present 
context makes use of the collinear subtraction structure 
already defined in the context of Deep Inelastic Scattering. 
Despite the fact the the full results make use of fracture functions 
and threfore a phenomenological modelling of the latter would be eventually 
required, the advatages reside in that fracture functions 
embodies the correct scale dependence through their own evolution equations.

\appendix
\section{Finite coefficients}
\label{finite_coef}
In this Appendix we present the results for the finite coefficient which appear in the previous sections.
The plus distribution are defined in the usual way:
\begin{equation}
\int_0^1 dw \frac{h(w)}{(1-w)_{+[0,\underline{1}]}} \equiv \int_0^1 dw \frac{h(w)-h(1)}{1-w}.
\end{equation}
The subtraction point is underlined. 
The coefficient functions in the target fragmentation region do coincide 
with the one found in the inclusive case and read:
\begin{eqnarray}
 && C_{q\bar{q}}\Big( w, \frac{\mu_F^2}{Q^2}\Big)=- 2 P_{qq}(w)
\ln\frac{\mu_F^2}{Q^2}+\widetilde{C}_{q\bar{q}}(w) \,,\nonumber\\
 && C_{qg}\Big( w, \frac{\mu_F^2}{Q^2}\Big) = - P_{qg}(w) \ln\frac{\mu_F^2}{Q^2} 
 + \widetilde{C}_{qg}(w)\,, 
\end{eqnarray}
where the scale independent coefficients $\widetilde{C}_{ij}$ are given by:
\begin{eqnarray}
&& \widetilde{C}_{q\bar{q}}(w)=C_F \Bigg[
4(1+w^2)\Bigg( \frac{\ln(1-w)}{1-w} \Bigg)_{+[0,1]} -2\frac{1+w^2}{1-w}\ln w
+\Big(\frac{2}{3}\pi^2-8\Big)\delta(1-w) \Bigg]\,,\nonumber\\
 && \widetilde{C}_{qg}(w)= T_R \Bigg[
\big(w^2+(1-w)^2\big)\ln \frac{(1-w)^2}{w}+\frac{1}{2}+3w-\frac{7}{2}w^2\,\Bigg]\,.
\end{eqnarray}
The polynomial term in $C_{qg}$ is slightly different from the one reported in Ref.~\cite{DYNLO}
because an additional term $(1-\epsilon)^{-1}$ is provided for matrix elements with a gluon in the 
initial state. This accounts for the correct  
gluon polarization in $n$-dimensions and it affects only the polynomial terms 
in the coefficient function. 
For the central term the subtraction is performed on the angular variable $y$.
The plus distributions in eq.~(\ref{yplus})
are defined by:
\begin{eqnarray}
\int_0^1 dy \frac{h(y)}{(1-y)_{+[0,\underline{1}]}} \equiv \int_0^1 dy \frac{h(y)-h(1)}{1-y}\,,\\
\int_0^1 dy \frac{h(y)}{(y)_{+[\underline{0},1]}} \equiv \int_0^1 dy \frac{h(y)-h(0)}{y}\,,
\end{eqnarray}
where $h(y)$ is a smooth test function. 
The coefficients $K_{ij}$ read:
\begin{eqnarray}
\label{Kcoeftilde}
K_{q\bar{q}}\Big(y,x_1,x_2,\tau,\frac{\mu_F^2}{Q^2}\Big)
&=&\widetilde{K}_{q\bar{q}}(y,x_1,x_2,\tau)+\nonumber\\
&&-\delta(y) \widehat{P}_{qq}(w) \frac{x_2}{x_1 x_2 -\tau}
\ln\frac{\mu_F^2}{Q^2}-\delta(1-y) \widehat{P}_{qq}(w) \frac{x_1}{x_1 x_2 -\tau}
\ln\frac{\mu_F^2}{Q^2}\,;\nonumber\\
K_{qg}\Big(y,x_1,x_2,\tau,\frac{\mu_F^2}{Q^2}\Big)
&=&\widetilde{K}_{qg}(y,x_1,x_2,\tau)-\delta(y) \frac{x_2}{x_1 x_2 -\tau} \widehat{P}_{qg}(w)
\ln\frac{\mu_F^2}{Q^2}\,.
\end{eqnarray}
By defining 
$b(x_1,x_2,y)=x_1y+x_2(1-y)$ and $a=x_1 x_2-\tau$, the scale-independent 
$\widetilde{K}_{ij}$ coefficients read:
\begin{eqnarray}
\label{Kcoef}
\widetilde{K}_{q\bar{q}}(y,x_1,x_2,\tau)&=&
C_F\Bigg\{\frac{1}{b}\Bigg(\frac{x_2}{x_1}(1-y)+\frac{2 \tau b^2}{a^2}\Bigg)\, 
\frac{1}{y_{+[\underline{0},1]}}+
\frac{1}{b}\Bigg(\frac{x_1}{x_2}y+\frac{2 \tau b^2}{a^2}\Bigg)
\frac{1}{(1-y)_{+[0,\underline{1}]}}+\nonumber\\  
&&+\,\delta(y)\Bigg[ \frac{1}{x_1} - 
\frac{1+w^2}{1-w}\frac{x_2}{a}
\ln\frac{x_2^2 \tau}{a^2} \Bigg]
+\delta(1-y)\Bigg[ \frac{1}{x_2} -
\frac{1+w^2}{1-w}\frac{x_1}{a}
\ln\frac{x_1^2 \tau}{a^2} \Bigg] \Bigg\}\,; \nonumber\\
\widetilde{K}_{qg}\Big(y,x_1,x_2,\tau)&=&
T_R\Bigg\{ \Bigg( \frac{x_2}{a}-\frac{2(1-y)\tau}{b x_1^2} \Bigg) 
\frac{1}{y_{+[\underline{0},1]}}+\frac{ay}{x_2 b^2}+\nonumber\\
 &&+\delta(y) \frac{x_2}{a} \Bigg[ 1- \Big[w^2+(1-w)^2\Big]
\Bigg( \ln\frac{\tau x_2^2}{a^2}+1\Bigg)\Bigg]\Bigg\}\,.
\end{eqnarray}
The coefficient function 
$K_{gq}$ can be obtained by exchanging $x_1\leftrightarrow x_2$ and $y \leftrightarrow 1-y$ in 
the expressions for $K_{qg}$. In $K_{q\bar{q}}$ one can note the appearance of poles in $(1-w)^{-1}$, 
where $w= \tau/(x_1 x_2)$. In this limit only soft parton emissions are allowed. 
This limit however is outside the integration region specified 
in the central contribution to the final result, eq.~(\ref{siDY_total_finite}),
due to the requirement $z >0$ embodied in the specific form of the integration boundaries $r_1$ and 
$r_2$ in eqs.~(\ref{r12}).

\end{document}